\documentclass[8.5pt,twoside,twocolumn]{article}
\usepackage[super,sort&compress,comma]{natbib}
\usepackage{mhchem}
\usepackage{amsmath}
\usepackage{times,mathptmx}
\usepackage{sectsty}
\usepackage{balance}

\usepackage{graphicx} 
\usepackage{lastpage}
\usepackage[format=plain,justification=raggedright,singlelinecheck=false,font=small,labelfont=bf,
labelsep=space]{caption}
\usepackage{fancyhdr}
\newcommand{\eq}[1]{(\ref{#1})}
\newcommand{\fig}[1]{Fig.\ref{#1}}

\newcommand{\be}{\begin{equation}}
\newcommand{\ee}{\end{equation}}

\newcommand\disp{\displaystyle}

\newcommand{\re}{\textrm{Re}\,}
\newcommand{\im}{\textrm{Im}\,}

\begin{document}

\thispagestyle{plain} \fancypagestyle{plain}{
\fancyhead[L]{\includegraphics[height=8pt]{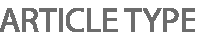}}
\fancyhead[C]{\hspace{-1cm}\includegraphics[height=20pt]{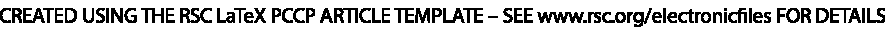}}
\fancyhead[R]{\includegraphics[height=10pt]{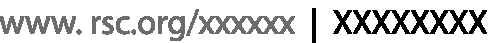}\vspace{-0.2cm}}
\renewcommand{\headrulewidth}{1pt}}
\renewcommand{\thefootnote}{\fnsymbol{footnote}}
\renewcommand\footnoterule{\vspace*{1pt}%
\hrule width 3.4in height 0.4pt \vspace*{5pt}} \setcounter{secnumdepth}{5}

\makeatletter
\def\subsubsection{\@startsection{subsubsection}{3}{10pt}{-1.25ex plus -1ex
minus -.1ex}{0ex plus 0ex}{\normalsize\bf}}
\def\paragraph{\@startsection{paragraph}{4}{10pt}{-1.25ex plus -1ex
minus -.1ex}{0ex plus 0ex}{\normalsize\textit}}
\renewcommand\@biblabel[1]{#1}
\renewcommand\@makefntext[1]%
{\noindent\makebox[0pt][r]{\@thefnmark\,}#1} \makeatother
\renewcommand{\figurename}{\small{Fig.}~}
\sectionfont{\large}
\subsectionfont{\normalsize}

\fancyfoot{} \fancyfoot[LO,RE]{\vspace{-7pt}\includegraphics[height=9pt]{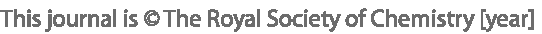}}
\fancyfoot[CO]{\vspace{-7.2pt}\hspace{12.2cm}\includegraphics{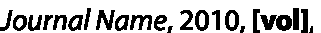}}
\fancyfoot[CE]{\vspace{-7.5pt}\hspace{-13.5cm}\includegraphics{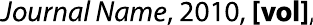}}
\fancyfoot[RO]{\footnotesize{\sffamily{1--\pageref{LastPage} ~\textbar  \hspace{2pt}\thepage}}}
\fancyfoot[LE]{\footnotesize{\sffamily{\thepage~\textbar\hspace{3.45cm} 1--\pageref{LastPage}}}}
\fancyhead{}
\renewcommand{\headrulewidth}{1pt}
\renewcommand{\footrulewidth}{1pt}
\setlength{\arrayrulewidth}{1pt} \setlength{\columnsep}{6.5mm} \setlength\bibsep{1pt}

\twocolumn[\begin{@twocolumnfalse}
\noindent\LARGE{\textbf{From geometric optics to plants: eikonal equation for buckling}}
\vspace{0.6cm}

\noindent\large{\textbf{Sergei Nechaev$^{\ast}$\textit{$^{a,b}$} and
Kirill Polovnikov,\textit{$^{c}$}}}\vspace{0.5cm}

\noindent\textit{\small{\textbf{Received Xth XXXXXXXXXX 20XX, Accepted Xth XXXXXXXXX 20XX\newline
First published on the web Xth XXXXXXXXXX 200X}}}

\noindent \textbf{\small{DOI: 10.1039/b000000x}}
\vspace{0.6cm}

\noindent \normalsize{Optimal embedding in the three-dimensional space of exponentially growing
squeezed surfaces, like plants leaves, or 2D colonies of exponentially reproducing cells, is
considered in the framework of conformal approach. It is shown that the boundary profile of a
growing tissue is described by the 2D eikonal equation, which provides the geometric optic
approximation for the wave front propagating in the media with inhomogeneous refraction
coefficient. The variety of optimal surfaces embedded in 3D is controlled by spatial dependence
of the refraction coefficient which, in turn, is dictated by the local growth protocol.}
\vspace{0.5cm}
\end{@twocolumnfalse}]

\section{Introduction}


\footnotetext{\textit{$^{a}$~J.-V. Poncelet Laboratory, CNRS, UMI 2615, 119002 Moscow, Russia;
E-mail: sergei.nechaev@gmail.com}}

\footnotetext{\textit{$^{b}$~P.N. Lebedev Physical Institute, RAS, 119991 Moscow, Russia}}

\footnotetext{\textit{$^{a}$~Physics Department, M.V. Lomonosov Moscow State University, 119992
Moscow, Russia}}


Variety of shapes of 2D growing tissues emerges due to the incompatibility of local internal
(differential) growth protocol with geometric constraints imposed by embedding of these tissues
into the space. For example, buckling of a lettuce leaf can be naively explained as a conflict
between natural growth due to the periphery cells division (typically, exponential), and growth of
circumference of a planar disc with gradually increasing radius. Due to a specific biological
mechanism which inhibits growth of the cell experiencing sufficient external pressure, the division
of inner cells is insignificant, while periphery cells have less steric restrictions and
proliferate easier. Thus, the division of border cells has the major impact on the instabilities in
the tissue. Such a differential growth induces an increasing strain in a tissue near its edge and
results in two complimentary possibilities: i) in-plane tissue compression and/or redistribution of
layer cells accompanied by the in-plane circumference instability, or ii) out-of-plane tissue
buckling with the formation of saddle-like surface regions. The latter is typical for various
undulant negatively curved shapes which are ubiquitous to many mild plants growing up in air or
water where the gravity is of sufficiently small matter \cite{kelp,swinney}.

A widely used energetic approach to growing patterns exploits a continuous formulation of the
differential growth and is based on a rivalry between bending and stretching energies of elastic
membranes \cite{hebrew,lewicka,audoly, gemmer, swinney, shape-of-leaf, muller,goriely}, reflecting
the choice between options (i) and (ii) above. For bending rigidity of a thin membrane, ${\cal
B}$, one has ${\cal B} \sim h^3$, while stretching rigidity, ${\cal S}$ behaves as ${\cal S}
\sim h$, where $h$ is the membrane thickness \cite{rayleigh}. Therefore, thin enough tissues, with
$h\ll 1$, prefer to bend, i.e. to be negatively curved under relatively small critical strain.

The latter allows one to eliminate the "stretching" regime from consideration, justifying the
geometric approach for infinitesimally thin membranes \cite{1,2,marder,marder-sharon, marder-pap,
gemmer} (see also \cite{3}). Here the determination of typical profiles of buckling surfaces relies
on an appropriate choice of metric tensor of the non-Euclidean space, and is realized via the
optimal embedding of the tissue with certain metrics into the 3D Euclidean space. It should be
mentioned, that the formation of wrinkles within this approach seems to be closely related to the
description of phyllotaxis via conformal methods \cite{levitov}.

In this letter we suggest a model of a hyperbolic infinitesimally thin tissue, whose periphery
cells divide freely with exponential rate, while division of inner cells is absolutely inhibited.
Two cases of proliferations, the one-dimensional (directed) and the uniform two-dimensional, are
considered. The selection of these two growth models is caused by the intention to describe
different symmetries inherent for plants at initial stages of growth. As long as the in-plane
deformations are not beneficial, as follows from the relationship between bending and stretching
rigidities, all the redundant material of fairly elastic tissue will buckle out. In order to take
into account the finite elasticity of growing tissue, resulting from the intrinsic discrete
properties of a material, we describe the tissue as a collection of glued elementary plaquettes
connected along the hyperbolic graph, $\gamma$. The discretization implies the presence of a
characteristic scale, of order of the elementary cell (plaquette) size, below which the tissue is
locally flat.

As we rely on the absence of in-plane deformations, this graph has to be isometrically embedded
into the 3D space. The desired smooth surface profile is obtained in two steps: i) isometric
mapping of the hyperbolic graph onto the flat domain (rectangular or circular) with hyperbolic
metrics, ii) subsequent restoring of the metrics into the 3D Euclidean space above the domain. We
demonstrate that such a procedure leads to the "optimal" buckling of the tissue and is described by
the eikonal equation for the profile, $f(x,y)$, of growing sample, which by definition, is a
variant of the Hamilton-Jacobi equation.

The paper is organized as follows. We introduce necessary definitions in the Section \ref{s:1}; the
model under consideration and the details of the conformal approach are provided in the Sections
\ref{s:2}, \ref{s:3} and Appendix; the samples of various typical shapes for
two-dimensional uniform and for one-dimensional directed growth, are presented in the Section
\ref{s:4}; finally, the results of the work are summarized in the Section \ref{s:5}, where we also
speculate about possible generalizations and rise open questions.

\section{Buckling of thin tissues in cylindric and planar geometries}

\subsection{Basic facts about the eikonal equation}
\label{s:1}

To make the content of the paper as self-contained as possible, it seems instructive to provide
some important definitions used at length of the paper. The key ingredient of our consideration is
the "eikonal" equation, which is the analogue of the Hamilton-Jacobi equation in geometric
optics. As we show below, the eikonal equation provides optimal embedding of an exponentially
growing surface into the 3D Euclidean plane. Meaning of the notion "optimal" has two different
connotations in our approach:

i) On one hand, from viewpoint of the Hamilton-Jacobi theory, the eikonal equation appears in the
minimization of the action $A=\int_\gamma L dt$ with some Lagrangian $L$. According to the Fermat
principle, the time of the ray propagation in the inhomogeneous media with the space-dependent
refraction coefficient, $n(x,y)$, should be minimal.

ii) On the other hand, the eikonal equation emerges in our work in a purely geometric setting
following directly from the conformal approach.

First attempts to formulate classical mechanics problems in geometric optics terms goes back to the
works of Klein \cite{klein} in 19th century. His ideas contributed to the corpuscular theory in a
short-wavelength regime, as long as the same mechanical formalism applied to massless particles,
was consistent with the wave approach. Later, in the context of general relativity, this approach
was renewed to treat gravitational field as an optic medium \cite{rumer}.

The Fermat principle states that the time $dt$ for a ray to propagate along a curve $\gamma$
between two closely located points $M({\bf x})$ and $N({\bf x}+d{\bf x})$ in an inhomogeneous
media, should be minimal. The total time $T$ can be written in the form $T=\frac{1}{c}\int_{M}^{N}
n(\mathbf{x}(s))ds$ where $n(\mathbf{x})=\frac{c}{\textrm{v}(\mathbf{x})}$ is the refraction
coefficient at the point $\mathbf{x}=\{x^i\}$ of a $D$-dimensional space $(i=1,...,D)$, $c$ and
$\textrm{v}(\mathbf{x})$ are correspondingly the light speeds in vacuum and in the media, and
$d|\mathbf{x}|=ds$ is the spatial increment along the ray. Following the optical-mechanical
analogy, according to which the action in mechanics corresponds to eikonal in optics, one can write
down the "optic length" or eikonal, $S=cT$ in Lagrangian terms: $S=\int_{M}^{N} L(\mathbf{x},
\dot{\mathbf{x}}) ds$ with the Lagrangian $L(\mathbf{x},\dot{\mathbf{x}})= n(\mathbf{x}(s))
\sqrt{\dot{\mathbf{x}}(s)\dot{\mathbf{x}}(s)}$, where $\dot{\mathbf{x}}^2= \sum_{i=1}^D
\big(\frac{dx^i}{ds}\big)^2$. We would like to mention here, that optical properties of the media
can be also treated in terms of induced Riemann metrics in vacuum:
\be
S=\int_{M}^{N}n(\mathbf{x}(s)) ds = \int_{M}^{N} \sqrt{\dot{\mathbf{x}} g(\mathbf{x})
\dot{\mathbf{x}}} ds
\label{eq:eikonal-metrics}
\ee
where $g_{ij} = n^2(\mathbf{x})\delta_{ij}$ stands for induced metrics components in isotropic media case.
Thus, from the geometrical point, the ray trajectory can be understood as a "minimal curve" in a
certain Riemann space. This representation suggests to consider optimal ray paths as geodesics in
the space with known metrics $g$.

Stationarity of optic length, $S$, i.e. $\delta S=0$, together with the condition
$|\dot{\mathbf{x}}|=1$, defines the Euler equation:
\be
\frac{d}{ds}\left(n(\mathbf{x})\frac{d\mathbf{x}}{ds}\right)= \nabla n(\mathbf{x})
\label{eq:euler}
\ee
from which one can directly proceed to the Huygens principle by integrating \eq{eq:euler} over $s$:
$\nabla S(\mathbf{x})=n(\mathbf{x})\frac{d\mathbf{x}}{ds}$. Squaring both sides of the latter
equation we end up with the eikonal equation:
\be
\left(\nabla S(\mathbf{x})\right)^2=n^2(\mathbf{x})
\label{eq:eikonal}
\ee
The eikonal equation Eq.\eq{eq:eikonal} has the same form as the Hamilton-Jacobi equation in
mechanics for action in the $D+1$-dimensional space, which in turn can be understood as the
relativistic equation for the light, propagating in the Riemannian space.

\subsection{The model: formalization of physical ideas}
\label{s:2}

In our work the eikonal equation arises in the differential growth problem in a purely geometric
setting. Consider a tissue, represented by a colony of cells, growing in space without any
geometric constraints. The local division protocol is prescribed by nature, being particularly
recorded in genes and is accompanied by their mutations \cite{genes}. The exponential cell
division is implied, as already mentioned above. To make our viewpoint more transparent, suppose
that all cells, represented by equilateral triangles, divide independently and their
proliferation is initiated by the first "protocell". Connecting the centers of neighboring
triangles by nodes, we rise a graph $\gamma$. The number of vertices, $P_{\gamma}(k)$, in the
generation $k$, grows exponentially with $k$: $P_{\gamma}(k) \sim c^k$ ($c>1$). It is known that
exponential graphs possess hyperbolic metrics, meaning that they can be isometrically (with
fixed branch lengths and angles between adjacent branches) embedded into a hyperbolic plane. Thus,
it is clear, that the corresponding surface, pulled on the isometry of such graph in the 3D
Euclidean space, should be negatively curved.

\begin{figure}[h]
\centering
\includegraphics[width=8cm]{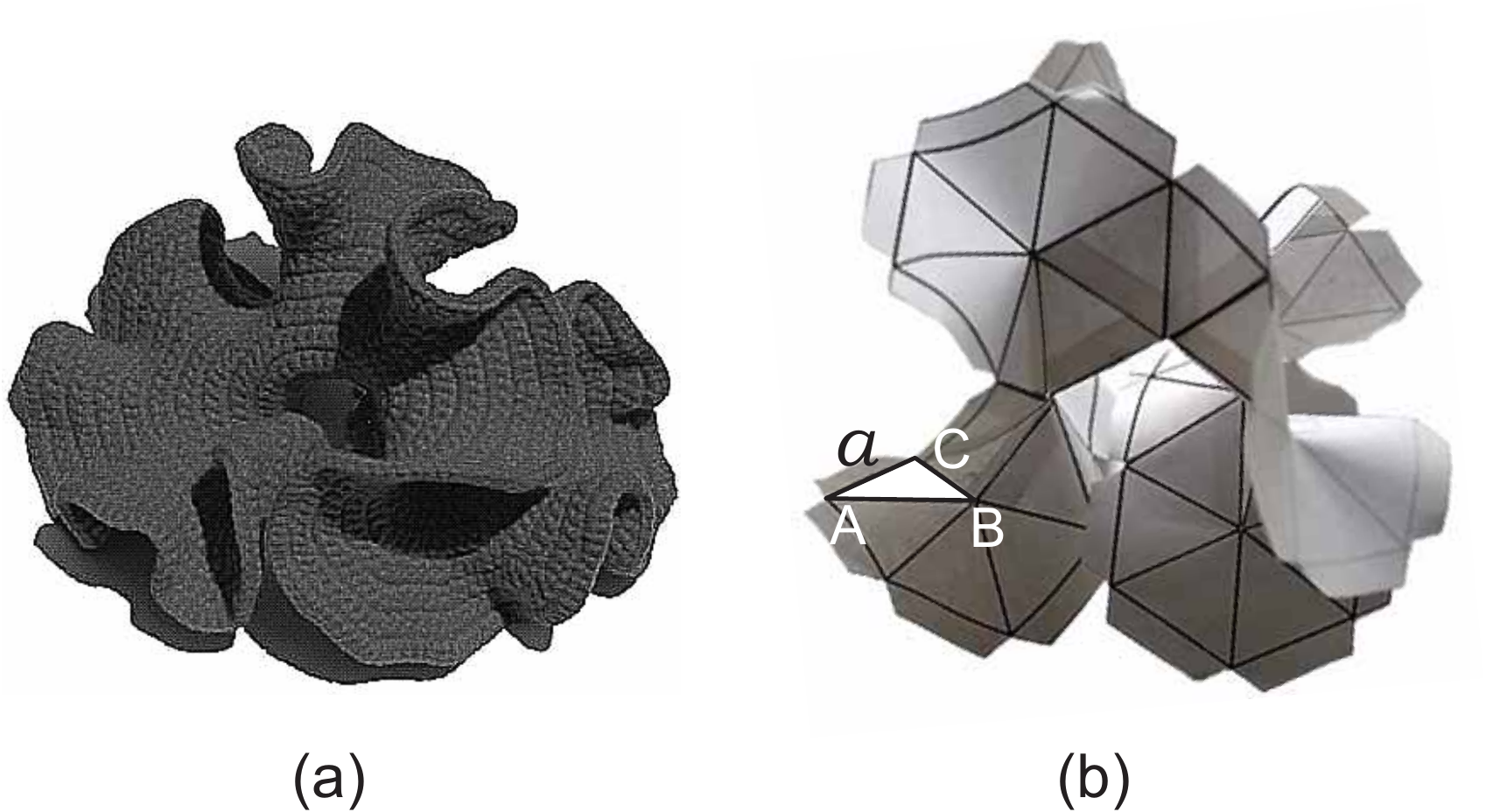}
\caption{(a) Hyperbolic surface obtained by spiral crocheting from the center; (b) Hyperbolic
piecewise surface constructed by joining 7 equilateral flat triangles (copies of the triangle
$ABC$) in each vertex. The triangle $ABC$ is lying in $z = x+iy$ plane in the 3D Euclidean space,
$|AB|=|AC|=|BC|=a$.}
\label{fig:01}
\end{figure}

To have a relevant image, suppose that we grow the surface by crocheting it spirally
starting from the center \cite{chroch}. Demanding two nearest neighboring circumference layers, $P(r)$ and
$P(r+\Delta r)$, to differ by a factor of $c$ (where $c=\mathrm{const}>1$), i.e., $P(r+\Delta
r)/P(r)=c$, we construct an exponentially growing (hyperbolic) surface -- see the \fig{fig:01}a
known as Amsler surface \cite{amsler}. The crocheted surface has well-posed properties on large
scales, but should be precisely described on the scale of order of the elementary cell. As we have
mentioned, the microscopic description is connected with the specific local growth protocol. The
simplest way to generate the discrete hyperbolic-like surface out of equilateral triangles,
consists in gluing 7 such triangles in each graph vertex and construct a piecewise surface, shown
in the \fig{fig:01}b. On the scale less than the elementary cell $ABC$ this surface is flat. Thus,
the size $a$ ($|AB|=|AC|=|BC|=a$) of the triangle $ABC$ stands for the rigidity parameter, playing
the role of a characteristic scale in our problem, below which no deviations from the Euclidean
metrics can be found. Later on we shall see that buckling of growing surface essentially depends
on this parameter.

\begin{figure}[h]
\centering
\includegraphics[width=8cm]{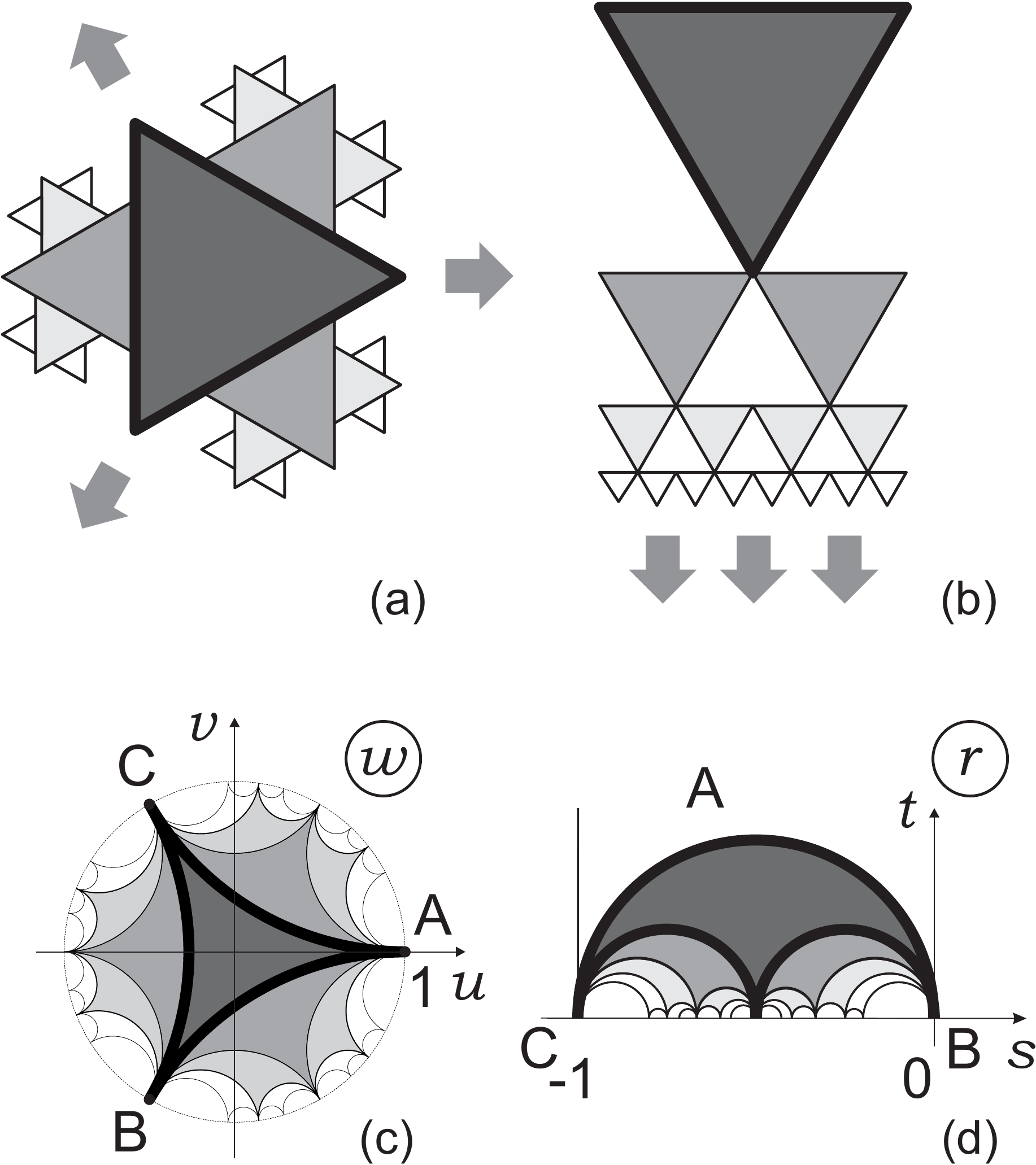}
\caption{(a) Uniform two-dimensional hyperbolic growth out of the unit domain in the plane;
(b) One-dimensional hyperbolic growth out of the linear segment; c) Tessellation of the hyperbolic
Poincar\'e disc by the images of flat Euclidean triangles; d) Tessellation of the domain in the
hyperbolic half-plane by the images of flat Euclidean triangles.}
\label{fig:02}
\end{figure}

We discuss buckling phenomena for two different growth symmetries shown schematically in the
\fig{fig:02}a-b: i) uniform two-dimensional division from the point-like source (\fig{fig:02}a), and ii) directed
one-dimensional growth from the linear segment (\fig{fig:02}b). In \fig{fig:02}a-b different
generations of cells are shown by the shades of gray. For convenience of perception, sizes of cells
in each new generation are decreasing in geometric progression, otherwise it would be impossible to
draw them in a 2D flat sheet of paper and the figure would be incomprehensible. In Figs.
\ref{fig:02}c,d we imitate the protocols of growth depicted above in Figs. \ref{fig:02}a,b by
embedding the exponentially growing structure in the corresponding plane domain equipped with the
hyperbolic metrics. The advantage of such embedding consists in the possibility to continue all
functions smoothly through the boundaries of elementary domains, that cover the whole plane without
gaps and intersections. Details of this construction and
its connection to the growth in the 3D Euclidean space are explained below.

It is known (see, for example \cite{marder}) that the optimal buckling surface is fully determined
by the metric tensor through the minimization of the discrete functional of special energetic form.
Namely, define the energy of a deformed thin membrane, having buckling profile $f(x,y)$ above
the domain, parameterized by $(x,y)$, as:
\be
E\{f(x,y)\} \sim \sum_{i,j} \left(\Big(f_{ij}\Big)^2-\sum_{\alpha,\beta}\Delta^{\alpha}_{ij}
g_{\alpha\beta} \Delta^{\beta}_{ij} \right)^2
\label{eq:E}
\ee
where $g_{\alpha\beta}$ is the induced metrics of the membrane, $f_{ij} \equiv |f(x_i, y_i) -
f(x_j, y_j)|$ is the distance between neighboring points and $\Delta_{ij}$ is the equilibrium
distance between them. The typical (optimal) shape $\bar{f}(x,y)$ is obtained by minimization of
\eq{eq:E} for any rigidity. However, the metric tensor, $g_{\alpha\beta}$ is a priori unknown
since its elements depend on specifics of the differential growth protocol, therefore some
plausible conjectures concerning its structure should be suggested. For example, in \cite{marder} a
directed growth of a tissue with one non-Euclidean metrics component, $g_{xx}(y)$, was considered.
The diagonal component $g_{xx}(y)$ was supposed to increase exponentially in the direction of the
growth, $y$, and crumpling of a leaf near its edge was finally established and analyzed.

\subsection{Conformal approach}
\label{s:3}

The preset rules of uniform exponential cells division determine the structure of the hyperbolic
graph, $\gamma$, while the infinitesimal membrane thickness allows for the isometrical embedding of
the graph $\gamma$ into the 3D space. We exploit conformal and metric relations between the surface
structure in the 3D space and the graph $\gamma$ embedded into the flat domain with the hyperbolic
metrics. The embedding procedure consists of a sequence of conformal transformations with a
constraint on area preservation of an elementary plaquette. This eventually yields the knowledge of
the Jacobian (the "coefficient of deformation"), $J(x,y)$, for the hyperbolic surface, which is
embedded into the 3D space via the orthogonal projection. Equipped by the key assumption, that a
smooth yet unknown surface $f(x,y)$ is \emph{function}, our procedure straightforwardly implies a
differential equation on the optimal surface. Note that a version of the Amsler surface cannot be
reconstructed in the same way since it is not a function above some planar domain.

\begin{figure}[ht]
\centering
\includegraphics[width=8cm]{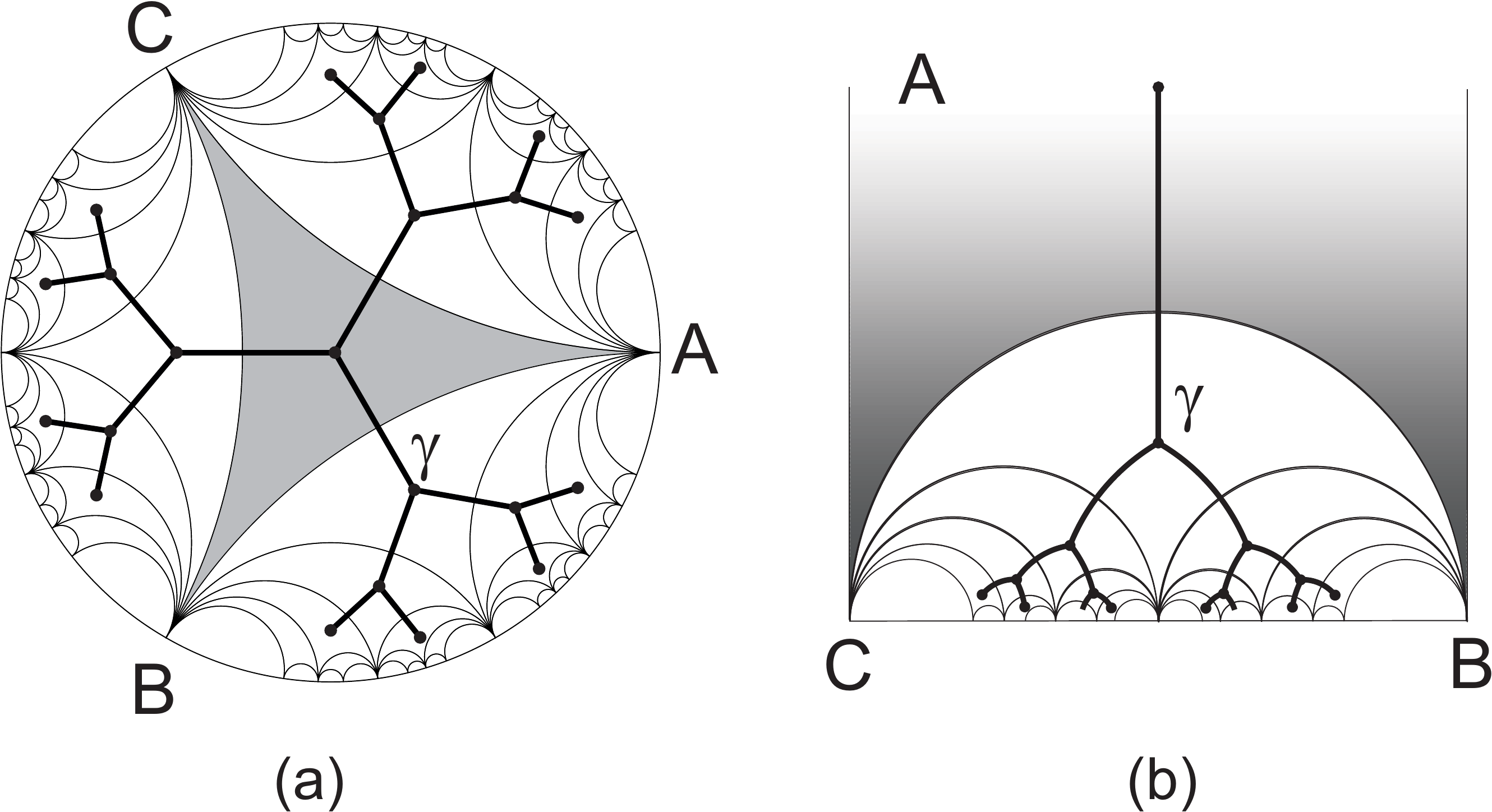}
\caption{Tessellation of the hyperbolic plane by the images of the
curvilinear triangle $ABC$: (a) for Poincar\'e disc; (b) for a strip of the upper half-plane. The
graph $\gamma$ connects the centers of images of $ABC$.}
\label{fig:03}
\end{figure}

To realize our construction explicitly, we first embed isometrically the graph $\gamma$: i) into
the Poincar\'e disk ($|w|<1$) for the model of uniform planar growth, and ii) into the strip of the
half-plane ($\im r > 0, -1<\re r<0$) for the model of one-dimensional growth. In the \fig{fig:03}
we have drawn the tessellation of the Poincar\'e disc and of the strip by equilateral curvilinear
triangles, which are obtained from the flat triangle $ABC$ of the hyperbolic surface (see
\fig{fig:01}b) by conformal mappings $z(w)$ and $z(r)$ discussed below. Note, that a conformal
mapping preserves the angles between adjacent branches of the graph. The graph
$\gamma$, shown in the \fig{fig:03}, connects the centers of the triangles and is
isometrically embedded into the corresponding hyperbolic domain. Besides, the areas of images
of the domain ABC are the same.

For the sake of definiteness consider the graph $\gamma$, isometrically embedded into the
hyperbolic disk, shown in the \fig{fig:03}a. Now, we would like to find the surface in the 3D
Euclidean space above the $w$-plane such that its Euclidean metrics coincides with the
non-Euclidean metrics in the disk. The Hilbert theorem \cite{hilbert} prohibits to do that for the
class of $C^2$-smooth surfaces. However, since we are interested in the isometric embedding of
piecewise surface consisting of glued triangles of fixed area, we can proceed with the standard
arguments of differential geometry \cite{diff-geom}. The metrics $ds^2$ of a 2D surface,
parameterized by $(u, v)$, is given by the coefficients
\be
E=\mathbf{r}_u^2, \quad F=\mathbf{r}_v^2, \quad G=(\mathbf{r}_u, \mathbf{r}_v)
\ee
of the first quadratic form of this surface:
\be
ds^2=E\, du^2 + 2F\, du dv + G\, dv^2
\ee
The surface area then reads $dS= \sqrt{EG-F^2}\, du dv$.

The area $S_{ABC}$ of the planar triangle $ABC$ on the plane $z = x + iy$ can be written as:
\be
S_{ABC}=\int\limits_{\triangle ABC} dx dy = \mathrm{const}
\label{eq:03}
\ee
where the integration is restricted by the boundary of the triangle. Since we aimed to conserve the
metrics, let us require that the area of the hyperbolic triangle $ABC$, after the conformal
mapping, is not changed and, therefore, it reads:
\be
S_{ABC} = \int\limits_{\triangle ABC} |J(z,w)| du dv; \quad J(z,w)=\left|\begin{array}{cc} \disp
\partial_u x & \disp \partial_u y \medskip  \\ \disp
\partial_v x & \disp \partial_v y
\end{array}\right|
\label{eq:04}
\ee
where $J(z,w)$ is the Jacobian of transition form $z$ to new coordinates, $w$. If $z(w)$ is
holomorphic function, the Cauchy-Riemann conditions allow to write
\be
J(w) = \left|\frac{dz(w)}{dw}\right|^2\equiv |z'(w)|^2.
\ee
From the other hand, we may treat the value of the Jacobian, $J(w)$, as a factor relating the
change of the surface element under transition to a new metrics, the co-called "coefficient of
deformation". As long as the metrics in the hyperbolic domain should reproduce the Euclidean metrics of the
smooth surface, $f(u, v)$, one should set $J = \sqrt {EG-F^2}$, where $E,G,F$ are the coefficients
of the first quadratic form of the surface $f$. Now, if $f(u,v)$ is function above $w$-plane, its
Jacobian adopts a simple form:
\be
J(u,v) = \sqrt{1+(\partial_u f)^2+(\partial_v f)^2}
\label{eq:06}
\ee

Making use of the polar coordinates in our
complex $w$-domain, $\{(\rho, \phi): u = \rho\cos\phi, v = \rho\sin\phi\}$, we eventually arrive
at nonlinear partial differential equation for surface profile $f(\rho,\phi)$ in the polar coordinates above $w$:
\be
\Big(\partial_\rho f(\rho,\phi)\Big)^2+\frac{1}{\rho^2}\Big(\partial_\phi
f(\rho,\phi)\Big)^2=|z'(w)|^4-1
\label{eq:08}
\ee

In the case of the hyperbolic strip domain, \fig{fig:03}b, the equation for the growth profile above the domain
can be written in local cartesian coordinates, $r = s + it$:

\be
\Big(\partial_s f(s, t)\Big)^2+\Big(\partial_t f(s,t)\Big)^2=|z'(r)|^4-1
\label{eq:07}
\ee

Note, that the inequalities $|z'(w)| > 1, |z'(r)| > 1$, following from \eq{eq:08}-\eq{eq:07}, determine
the local condition of existence of non-zero real solution and, as we discuss below, can be interpreted as the
presence of a finite scale surface rigidity.

To establish a bridge between optic and growth problems, let us mention that, say,
equation \eq{eq:08}, coincides with the two-dimensional eikonal equation \eq{eq:eikonal} for the
wavefront, $S(w)$, describing the light propagating according the Huygens principle in the
unit disk with the refraction coefficient
\be
n(w) = \sqrt{|z'(w)|^4-1}
\label{eq:n}
\ee

\begin{figure}[ht]
\centering
\includegraphics[width=6cm]{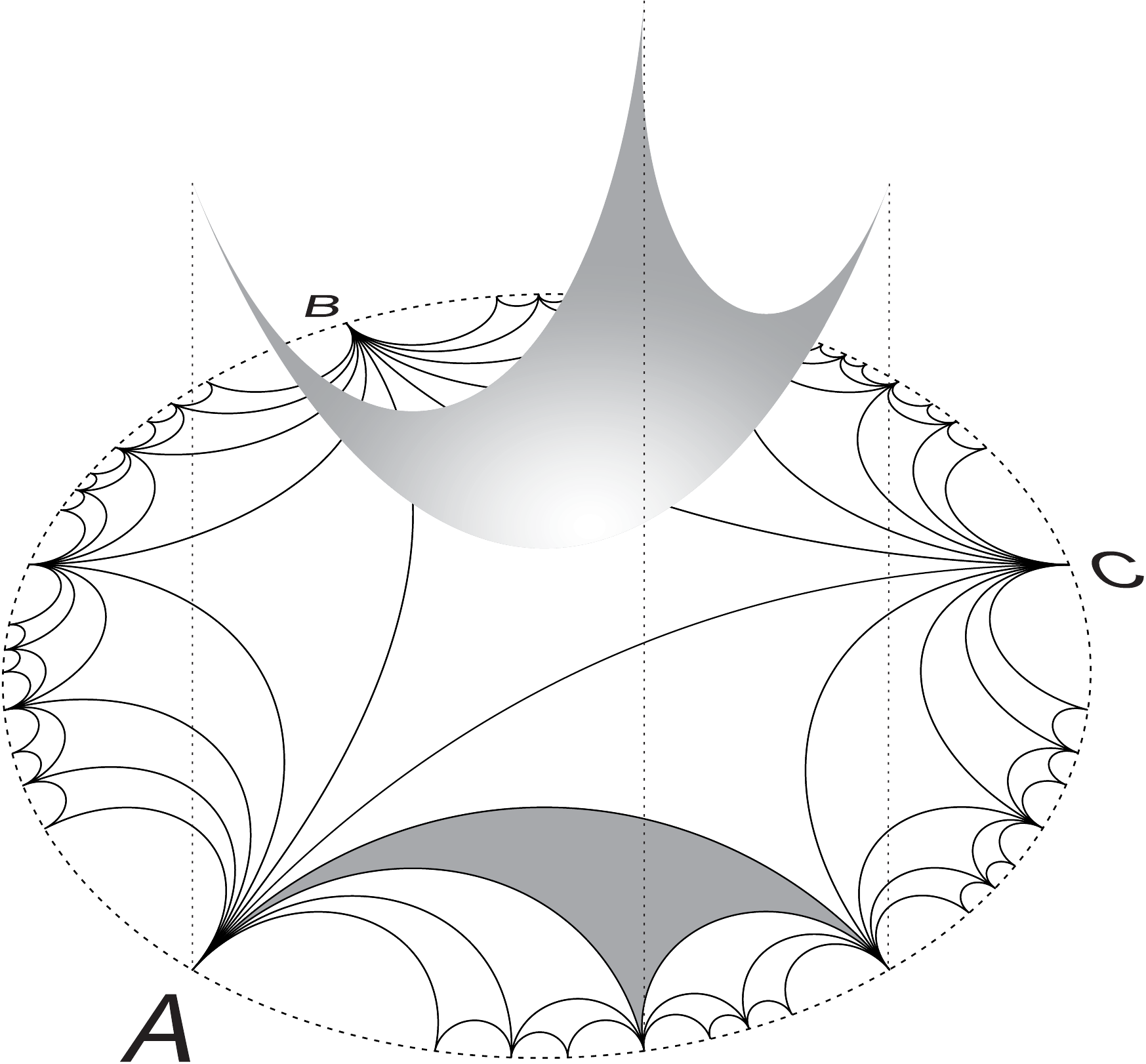}
\caption{Orthogonal projection above the Poincar\'e disc: area of the curvilinear triangle in Euclidean
space coincides with the area of the triangle in hyperbolic metrics in Poincar\'e disc.}
\label{fig:04}
\end{figure}

We can construct the conformal mappings $z(r)$ and $z(w)$ of the flat equilateral triangle $ABC$ in
the Euclidean complex plane $z=x+iy$ onto the circular triangle $ABC$ in the complex domains $r = s
+ it$ and $w=\rho(\cos\phi+i\sin\phi)$ correspondingly. The absolute value of the Gaussian
curvature is controlled by the number, $V$, of equilateral triangles glued in one vertex: the
surface is hyperbolic only for $V>6$. The surfaces with any $V>6$ have qualitatively similar
behavior, however the simplest case for analytical treatment corresponds to $V=\infty$, when the
dual graph $\gamma$ is loopless. The details of the conformal mapping of the flat triangle with
side $a$ to the triangle with angles $\{0,0,0\}$ in the unit strip $r$ are given in the Appendix.
The Jacobian $J(z(r))$ of conformal mapping $z \to r$ reads:
\be
J(r)=|z'(r)|^2 =\frac{h^2}{a^2}|\eta(r)|^8
\label{eq:ded1}
\ee
and the Jacobian of the mapping $z \to w$, is written through the function $r(w)$ that conformally
maps the triangle from the strip onto the Pincar\'e disk:
\be
J(w)=|z'(w)|^2 =\frac{3h^2}{a^2} \frac{|\eta(r(w))|^8}{|1-w|^4}
\label{eq:ded2}
\ee
where
\be
r(w) = e^{-i\pi/3}\frac{e^{2i\pi/3}-w}{1-w}-1; \quad h= \left(\frac{16}{\pi}\right)^{1/3}
\frac{\Gamma(\frac{2}{3})}{\Gamma^2(\frac{1}{3})} \approx 0.325
\ee
In both cases \eq{eq:ded1} and \eq{eq:ded2}, the function in the right-side of the equation is the
Dedekind $\eta$-function \cite{jacobi}:
\be
\eta(w)=e^{\pi i w/12}\prod_{n=0}^{\infty}(1-e^{2\pi i n w})
\label{eq:dedeta}
\ee

\section{Results and their interpretation}
\label{s:4}

The eikonal equation, \eq{eq:eikonal}, with \emph{constant} refraction index, $n$, corresponds to
optically homogeneous 2D domain, in which the light propagates along straight lines in Euclidean
metrics. From the other hand, in this case the eikonal equation yields the action surface with zero
Gaussian curvature: a conical surface above the disk, $S(\rho, \phi) \sim \rho$, for the uniform 2D
growth and a plane above the strip, $S(s, t) \sim t$, for the directed growth. Note, that at least
one family of geodesics of these surfaces consists of lines that are projected to the light
propagation paths in the underlying domain. We will show below that the geodesics of the eikonal
surface conserve this property even when the media becomes optically inhomogeneous.

For growth, the constant refraction index corresponds to an isometry of a planar growing surface
and absence of buckling. The conformal transformation, that results in the corresponding "coefficient
of deformation", $J^2(u, v) = n^2(u, v) + 1$, is uniformly compressive and the tissue remains
everywhere flat. Thus, it becomes clear, why the essential condition for buckling to appear is the
\emph{differential growth}, i.e. the spatial dependence of local rules of cells division.

We solve \eq{eq:08} and \eq{eq:07} numerically with the Jacobian, corresponding to exponentially
growing circumference, \eq{eq:ded1}-\eq{eq:ded2}, for different parameters $a$. We have chosen the
Dirichlet initial conditions along the line (for directed growth above the strip) and along the
circle of some small enough radius (for uniform 2D growth above the disk). The right-hand side of the
eikonal equations for the specific growth protocol is smooth and nearly constant up some radius and
then becomes more and more rugged. The constant plateau in vicinity of initial stages of growth is
related with the fact, that exponentially dividing cells can be organized in a Euclidean plane
up to some finite generations of growth. However, as the cells proliferate further, the isometry of
their mutual disposition becomes incompatible with the Euclidean geometry and buckling of the
tissue is observed. Note, that the Jacobian is angular-dependent, that is the artefact of chosen
triangular symmetry for the cells in our model. The existence of real solution, $\bar f(u, v)$ of
the eikonal equation is related to the sign of its right-hand side and is controlled by the parameter
$a$, while the complex solution $f(u, v) = f_R(u, v) + if_I(u, v)$ can be found for every $a$.

\begin{figure}[ht]
\centering
\includegraphics[width=8cm]{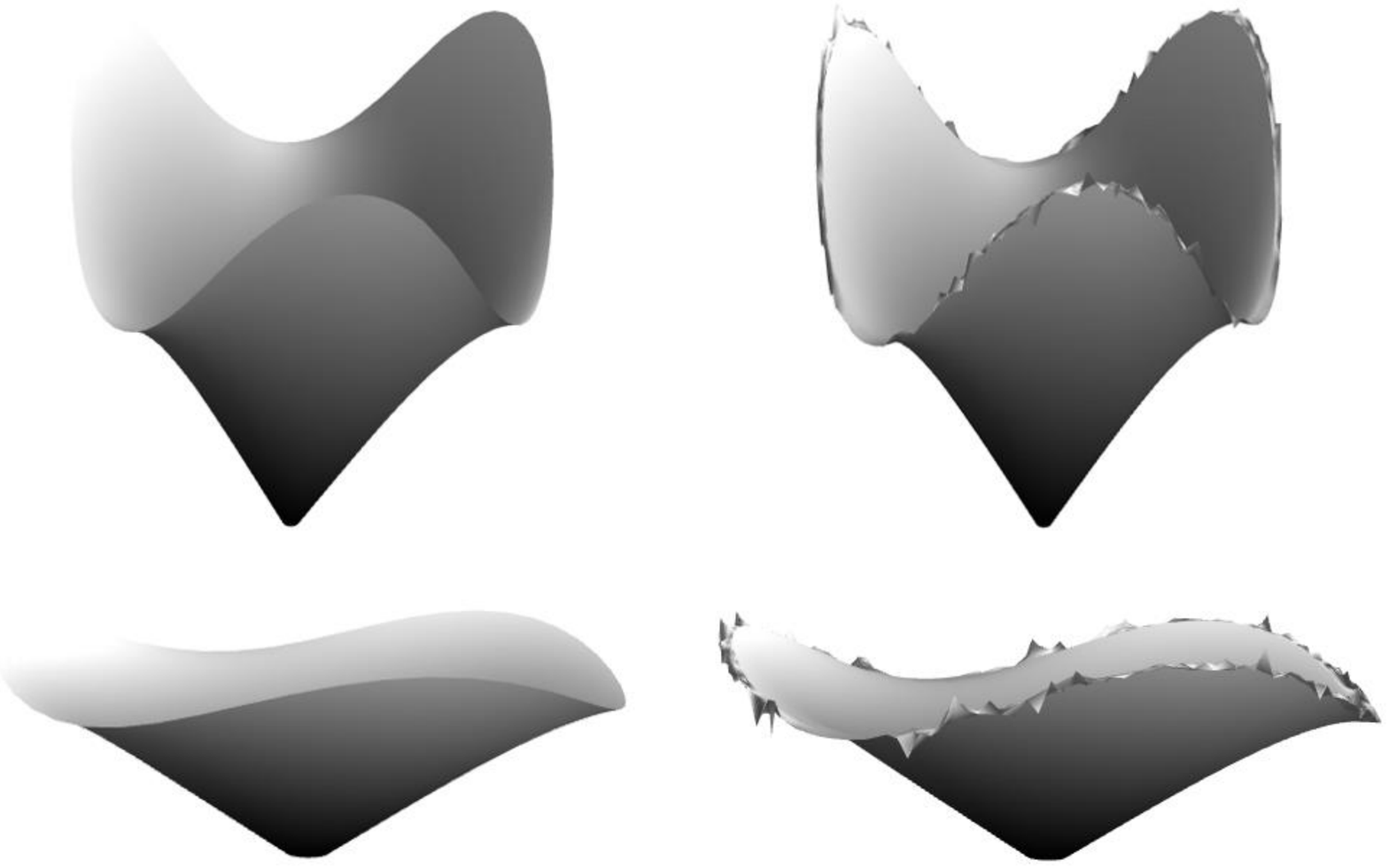}
\caption{The bluebell phase, obtained by numerical solution of \eq{eq:08} for two flexible tissues:
$a = 0.07$ (first row) and $a = 0.14$ (second row). Figures on the right show appearance of buckling
instabilities at the edge with growth.}
\label{fig:bluebell}
\end{figure}

First, we consider the 2D growth above the Poincar\'e domain, starting our numerics from low
enough values of $a$, for which the right-hand side of the eikonal equation, \eq{eq:08}, is strictly
positive on the plateau around the source of growth. Physically that means flexible enough tissues,
since, by construction, we require $a$ to be a scale on which the triangulated tissue does not
violate flat geometry. The real solution $\bar f(u,v)$ for these parameters exists up to late
stages of growth, see \fig{fig:bluebell} left. Note,
that a conical solution at early stages of growth is related to the plateau in the Jacobain and, as
it was discussed above, corresponds to the regime when cells can find place on the surface without
violation flat geometry. From the geometric optics point of view, that corresponds to constant
refraction index and straight Fermat geodesic paths in the underlying 2D domain. We show in the
\fig{fig:bluebell} that under increasing of $a$ the initial area of conical behavior is shrinking,
since the critical generation, at which the first buckling mode appears, is lower for larger cells.
In course of growth, the surface is getting negatively curved for some angular directions,
consistent with chosen triangular symmetry. It is found reminiscent of the shape of bluebells and,
in general, many sorts of flowers.

At late stages of growth, as we approach the boundary of the Poincar\'e disk, $\rho \to 1$ at some
fixed value of $\phi$, corresponding values of the right hand side of \eq{eq:08} become negative,
leading to the complex solution of the eikonal equation. Fortunately, we may infer some useful
information from the holomorphic properties of the eikonal equation in this regime, not too close
to the boundary of the disc. Applying the Cauchy-Riemann conditions to the solution of the eikonal
equation, $f$, we have: $\partial_u f_R = \partial_v f_I$ and $\partial_v f_R = -\partial_u f_I$.
Thus, the function $\bar f$ can be analytically continued in the vicinity of points along the curve
$\Gamma$ in the $(uv)$ plane, at which the right hand side of the eikonal equation nullifies. Moreover,
using this property, one can show, that the absolute value of the complex solution in the vicinity
of $\Gamma$ smoothly transfers to the real-valued solution, as one approaches the $\Gamma$ curve:
\be
\begin{array}{rll}
\disp \lim_{(u, v) \to \Gamma} \left(\nabla |f(u, v)|\right)^2 &=& \left(\nabla f_R(u,
v)\right)^2|_{\Gamma} \equiv \left(\nabla \bar f(u, v) \right)^2 \medskip \\
\disp |f(u, v)| &=& \sqrt{f_R^2(u, v) + f_I^2(u, v)}
\end{array}
\label{mod}
\ee

\begin{figure}[ht]
\centering
\includegraphics[width=8cm]{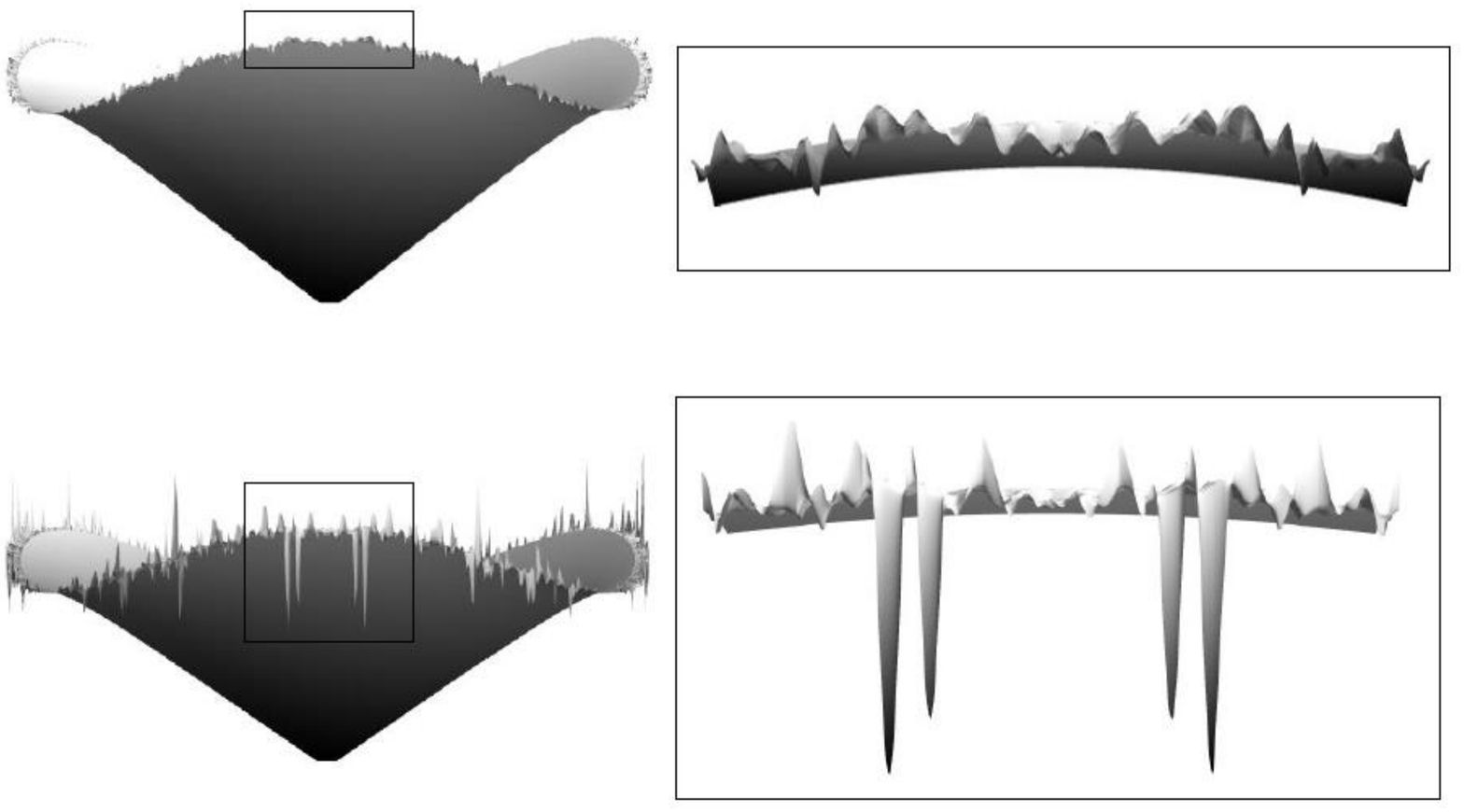}
\caption{Development of buckling instabilities at the edge of flower for rigidity parameter $a = 0.14$.
The right figures show hierarchical organization of the flower's circumference in detail.}
\label{fig:s-s}
\end{figure}

The non-existence of real solutions of the eikonal equation at late stages is a direct consequence
of the presence of finite bending scale, on which the tissue is locally flat. As it was
mentioned above and is shown in the \fig{fig:bluebell}, low values of $a$ lead to elongated conical
regime. Since $a$ stands for the scale on which the circumference length of the tissue doubles, in the $a \to 0$ limit the
real solution exists everywhere inside the disk, but it is everywhere flat (conical). Hopefully, the analytic continuation allows one
to investigate buckling for negative values of $n^2(u,v) = J^2(u, v)-1$ by taking the absolute
value of the solution, at least not far away from the zero-curve $\Gamma$. In this regime buckling
instabilities on the circumference of the bluebell arise. In the \fig{fig:s-s} we show
proliferation of buckling near the critical point. First, the evolution of buckling instabilities at the edge
can be understood as a subsequent doubling of peaks and saddles along the direction of growth. Then some hierarchy in peaks size
is seen. We note, that this hierarchical organization is a natural
result due to the theoretic-number properties of the Dedekind $\eta$-function. Though it is known, that in real plants and
flowers buckling instabilities do not proliferate profoundly, since the
division process is getting limited at late stages of growth, the formal continuation of the eikonal
equation beyond $\Gamma$ predicts a self-similar buckling profile at the circumference of growing tissues.

\begin{figure}[ht]
\centering
\includegraphics[width=8cm]{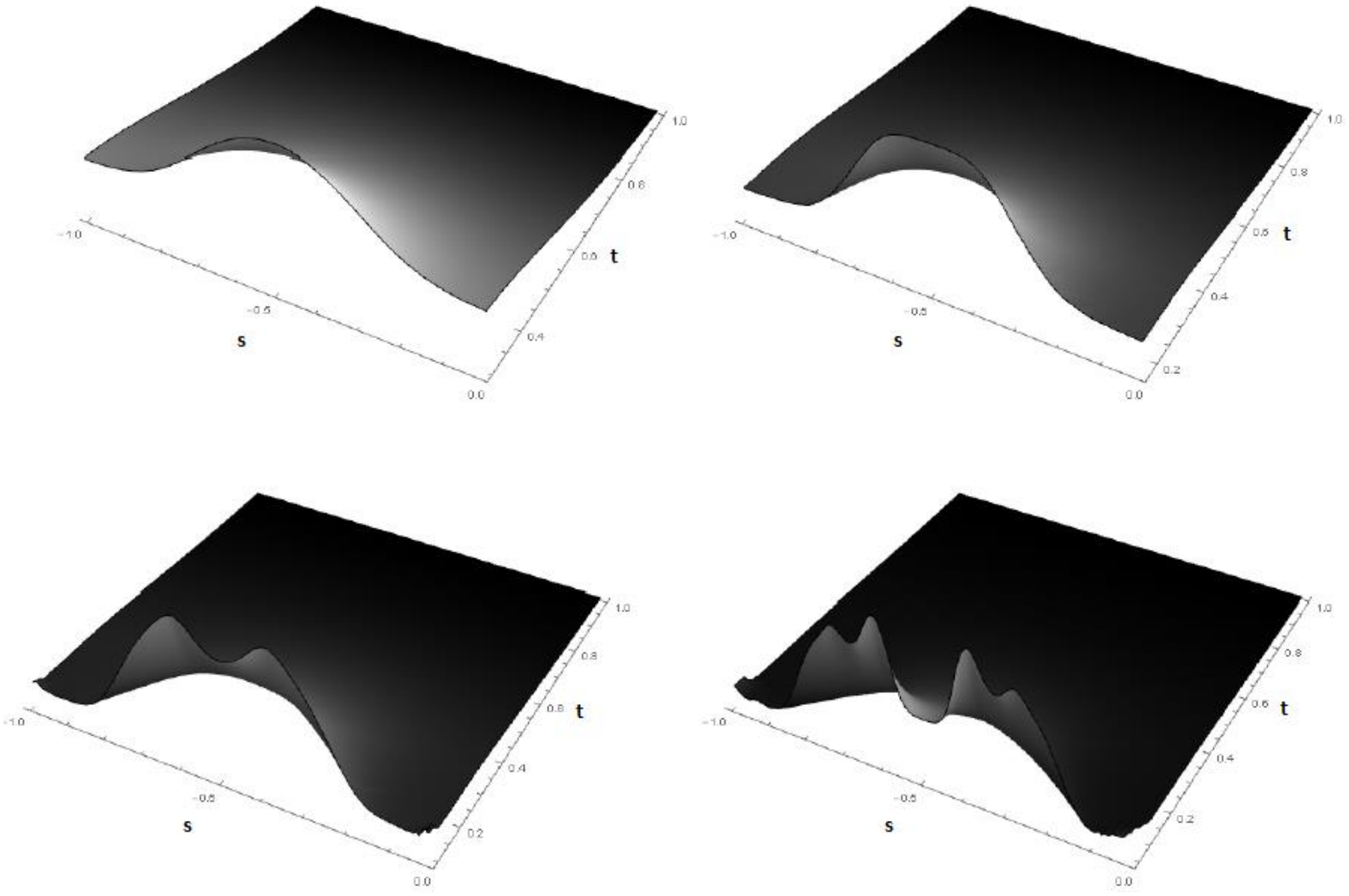}
\caption{Numerical solutions of \eq{eq:07} for the directed growth.
Figures show enhancing of buckling at the edge.}
\label{fig:direct_sol}
\end{figure}

Now we pay attention to the directed growth above the half-plane domain. Here we solve the equation
\eq{eq:07} with the Dirichlet boundary conditions, set along the line $t = 1$, and the tissue is
growing towards the boundary $t = 0$ in the upper halfplane $\im r>0$. At low stages of growth
the solution is flat until the first buckling mode appear, \fig{fig:direct_sol}. The subsequent
growth is described by taking the absolute value of the solution, since no real solution exists
anymore. As in the former case, the behavior is controlled by the value of $a$.

When the growth approaches the boundary, the edge of the tissue becomes more and more wrinkled.
Emergence of new buckling modes is the consequence of the Dedekind $\eta$-function properties: doubling
of parental peaks at the course of growth. Under the energetic approach for a leaf, very similar fractal structures
can be inferred from the interplay between
stretching and bending energies in the limit of extremely thin membranes: while the cell density
(and the corresponding strain, $\sigma$) on the periphery increases, the newly generating
wavelengths decrease, $\lambda\sim \sigma^{-1/4}$, \cite{mahadevan}.

Increasing the size $a$ of the elementary flat triangle domain, we figure out, that for some
critical value, $a_{cr}$, the starting plateau of the corresponding Jacobian crosses the zero level
and becomes negative. Our model implies no solutions for such stiff tissues. This limitation is
quite natural since we do not consider in-plane deformations of the tissue. In reality, for $a
> a_{cr}$ the tissue is so stiff, that it turns beneficial to be squeezed in-plane rather than to
buckle out. One may conjecture that $a$ is the analogue of the Young modulus, $E$, that is
known to regulate the rigidity of the tissue in the energetic approach, along with the thickness,
$h$, and the Poisson modulus, $\mu$, in their certain combination, known as bending stiffness, $D =
\frac{Eh^3}{12(1-\mu^2)}$.

\begin{figure}[ht]
\centering
\includegraphics[width=8cm]{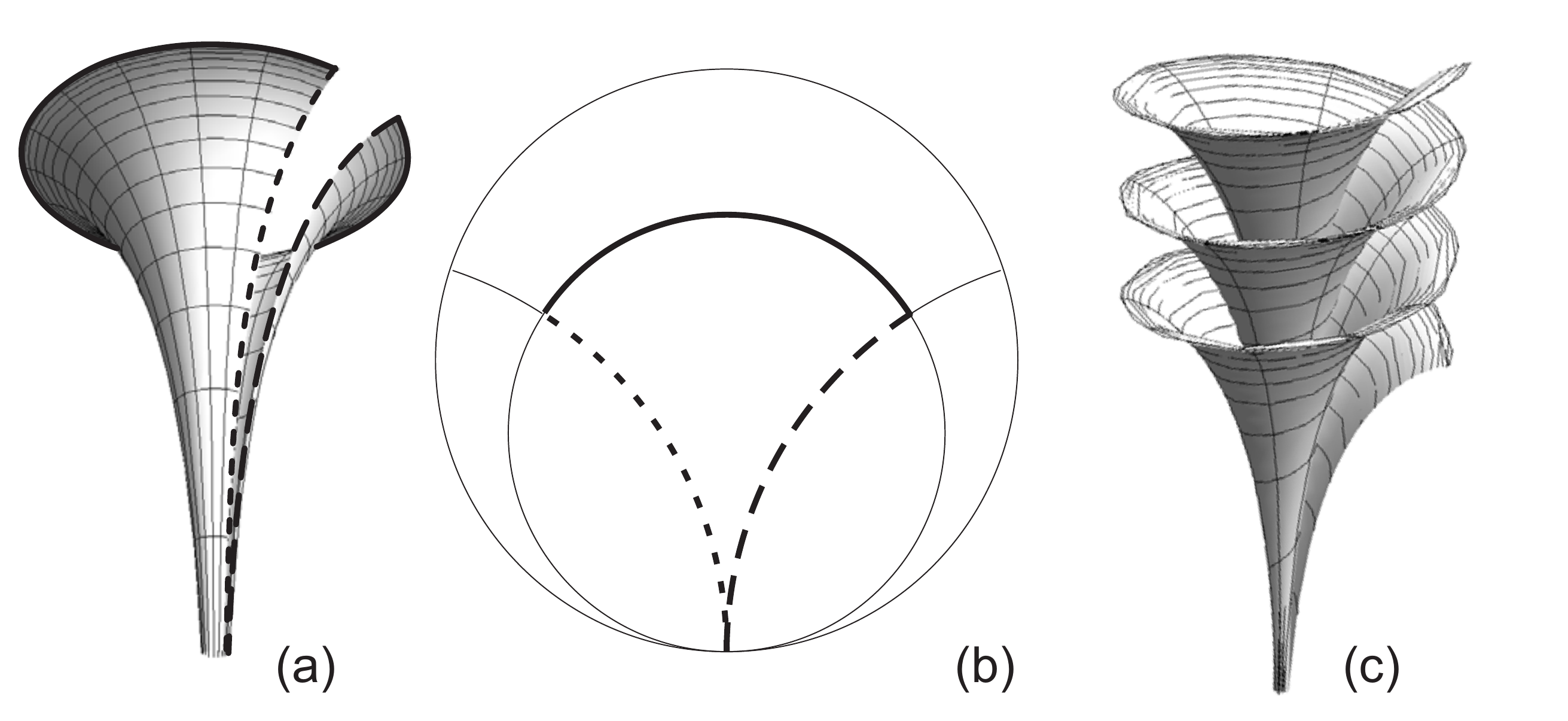}
\caption{(a)-(b) Pseudosphere and correspondence of boundaries on the Poincar\'e disc; (c) Dini
surface.}
\label{fig:pseudo}
\end{figure}

It is worth mentioning that at first stages of growth, until the instabilities at the circumference have not yet appeared,
at certain angles (triangle-like cells) the surface bends similar to the Beltrami's pseudosphere, that
has a constant negative curvature at every point of the surface, compare \fig{fig:bluebell} and \fig{fig:pseudo}.
The similarity is even more striking for very low $a$, when the triangulating parameter is fairly small.
It is known that the pseudosphere locally realizes the Lobachevsky geometry and can be isometrically mapped onto \emph{the finite
part} of the half-plane or of the Poincar\'e disk, \fig{fig:pseudo}a-b. According to the Hilbert theorem, \cite{hilbert}, no full
isometric embedding of the Poincar\'e disk into the 3D space exists. Thus, in order to organize
itself in the 3D space, the plant grows by the cascades of pseudospheres, resembling peaks and saddles,
that is an alternative view on essence of buckling.

Interestingly, some flowers, such as calla lilies, initially grow psuedospherically, but then
crack at some stage of growth and start twisting around in a helix. Apparently, this is another
route of dynamic organization of non-Euclidean isometry in the Euclidean space. The Dini's surface,
\fig{fig:pseudo}c is known in differential geometry as a surface of constant negative curvature and,
in comparison with the Beltrami's pseudosphere, is infinite. The problem of sudden cracking of the lilies
seems to be purely biological, but as soon as the crack appeared, the flower may relief the stresses caused by subsequent
differential growth through twisting its petals in the Dini's fashion.

Turn now to the eikonal interpretation of buckling. For the sake of simplicity,
we will proceed here in the cartesian coordinates. Seeking the solution of \eq{eq:eikonal} and
\eq{eq:07} in the implicit form $H(\mathbf{x})\equiv H(x^0,x^1,x^2) = H^0$ with $x^0= i f$,
$x^1=u$, $x^2=v$, we can rewrite \eq{eq:07} as:
\be
g^{ij}\frac{\partial H(\mathbf{x})}{\partial x^i}\frac{\partial H(\mathbf{x})}{\partial x^j}= 0;
\quad g^{ik} = \left(\begin{matrix} n^2 & 0 & 0 \\ 0 & 1 & 0 \\ 0 & 0 & 1 \end{matrix}\right)
\label{eq:rel}
\ee
Eq.\eq{eq:rel} reveals the relativistic nature of the eikonal equation \cite{landau} and describes
the propagation of light in a (2+1)D space-time in the gravitational field with induced metrics
$g$ defined by the metric tensor $g^{ik}$, where $n\equiv n(x^1,x^2)$, speed of light put $c=1$.
Having $g$, one can reconstruct geodesics that define the paths of the light propagation in our
space-time. The parameterized geodesics family, $x^{\lambda}(\tau)$, where $\lambda = 0, 1, 2$, can
be found from the equation:
\be
\frac{d^2 x^{\lambda}}{d \tau^2} + \Gamma^{\lambda}_{ij} \frac{dx^i}{d \tau} \frac{dx^j}{d \tau}=0
\label{geo}
\ee
where
\be
\Gamma^{i}_{kl} = \frac{1}{2} g^{im} \left(\frac{\partial g_{mk}}{\partial x^l} + \frac{\partial
g_{ml}}{\partial x^k} - \frac{\partial g_{kl}}{\partial x^m} \right)
\ee
are Christoffel symbols and $g_{ij}$ is the covariant form of the metrics ($g_{ij}g^{jk} =
\delta^k_i$). Calculating the symbols for the specific metrics \eq{eq:rel}, we end up with the set
of equations for the geodesics in a parametric form:
\be
\left\{\begin{array}{l}
\disp u_{\tau \tau} - \frac{1}{n^3} \frac{\partial n}{\partial u} f_{\tau}^2 = 0, \medskip \\
\disp v_{\tau \tau} - \frac{1}{n^3} \frac{\partial n}{\partial v} f_{\tau}^2 = 0, \medskip \\
\disp f_{\tau \tau} - \frac{2}{n}\frac{d n}{d \tau} f_{\tau} = 0
\end{array} \right.
\label{traj}
\ee
From the first two lines of \eq{traj}, one gets $\frac{u_{\tau \tau}}{v_{\tau \tau}} =
\frac{\partial n}{\partial u} \left(\frac{\partial n}{\partial v}\right)^{-1}$. Note, that the same relation
follows directly from \eq{eq:euler}, if the planar domain is parameterized by the same coordinates $\mathbf x = \mathbf x(u, v)$.
Thus, one may conclude, that the projections of the geodesics from the (2+1)D space-time onto the $(uv)$-plane coincide with
light trajectories in the flat domain with refraction coefficient $n(u,v)$.

\section{Conclusion and conjectures}
\label{s:5}

In this paper we discussed the optimal buckling profile formation of growing two-dimensional tissue
evoked by the exponential cell division from the point-like source and from the linear segment.
Such processes imply excess material generation enforcing the tissue to wrinkle as it approaches
the domain boundary. Resulting optimal hyperbolic surface is described by the eikonal equation for
the two-dimensional profile, and allows for simple geometric optics analogy. It is shown that the
surface height above the domain mimics the eikonal (action) surface of a particle moving in the 2D media with
certain refraction index, $n$, which, in turn, is linked to microscopic rules of elementary cell
division and symmetry of the plant. The projected geodesics of this "minimal" optimal surface coincide
with Fermat paths in the 2D media, which is the intrinsic feature of the eikonal equation. This result
suggests an idea to treat the growth process itself as a propagation of the wavefronts in the media
with certain metrics.

We have derived the metrics of the growing plant's surface from microscopic rules of cells division and
have shown that the solution of the eikonal equation describes buckling of tissues of different
rigidities. Our results, being purely geometric, rhyme well with a number of energetic approaches
to buckling of thin membranes, where the stiffness is controlled by the effective bending
rigidity. We show that presence of a finite scale on which the tissue remains flat, results in
negatively curved growing surfaces and the eikonal equation implies absence of real solution at
late stages of the growth. Though, an analytical continuation can be constructed and erratic
self-similar patterns along the circumference can be obtained. In reality high energetic costs for
the profound cell division after bifurcation point would prohibit infinite growth and intense buckling.

Recall, that the right-hand side of the eikonal equation mimics the squared refraction index, \eq{eq:n},
if buckling is interpreted as wavefront propagation in geometric optics. At length of our work it was pointed out,
that for the differential growth problem, negative square of refraction index leads to complex solution
for $f$. Does complex solution have any physical meaning for growth? We can provide the following speculation. The
complex solution appears for the late stages of growth when the finite bending scale of the tissue prohibits
formation of very low-wavelength buckling modes. Since in this regime the tissue
would experience in-plane deformations, one may improve the geometric model by letting
branches to accumulate the "potential energy". Thereby, the analogy between optics and differential growth
can be advanced by noting that the negative squared refraction index means absorbtion properties of the media. The
propagating wavefront of a moving particle, dissipates the energy in areas where the refraction index is complex-valued.
In the differential growth the proliferation of buckling modes may be limited by the energy losses at branches, that would
suppress buckling.

The challenging question concerns the possibility to extend our approach to the growth of
three-dimensional objects, for example, of a ball that size $R$ grows faster than $R^2$. In this
case, the redundant material can provoke the surface instabilities. We conjecture that some analogy
between the boundary growth and optic wavefronts survives in this case as well.

Authors are grateful to M. Tamm, A. Grosberg, M. Lenz, L. Mirny and L. Truskinovsly for valuable discussions
of various aspects of the work and to A. Orlov for invaluable help in numerical solution of the
eikonal equation. The work is partially supported by the IRSES DIONICOS and RFBR 16-02-00252A
grants.

\begin{appendix}

\section*{Appendix: Conformal transformation of the flat triangle to the Poincare domain}

\begin{figure}[ht]
\centering
\includegraphics[width=8cm]{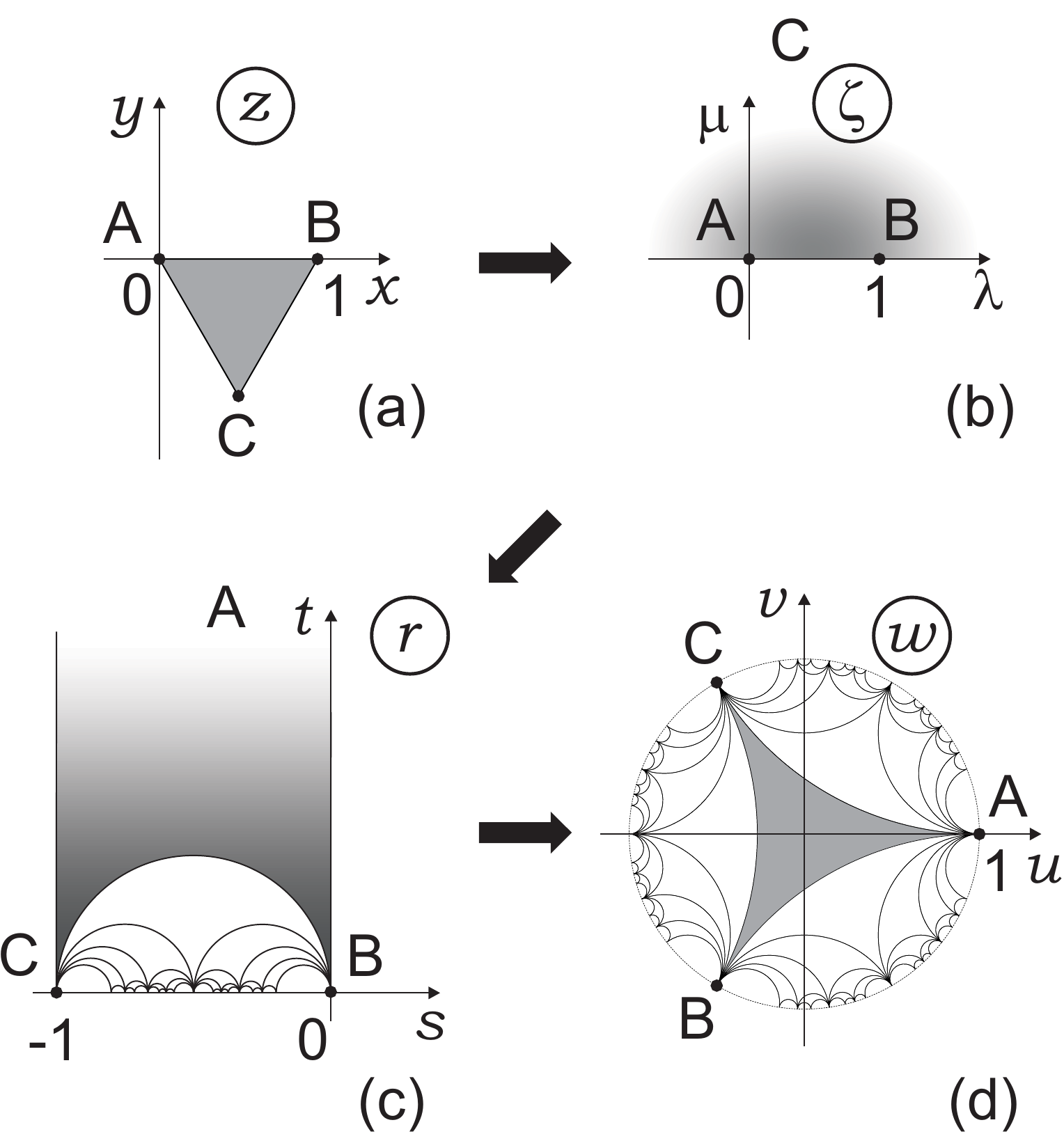}
\caption{Conformal mapping $z(w)$ is realized as a composition of three mappings: $z(\zeta)$
[(a)--(b)], $\zeta(r)$ [(b)--(c)], and $r(w)$ [(c)--(d)]. Finally we have $z(\zeta(r(w)))$.}
\label{fig:suppl01}
\end{figure}

The conformal mapping $z(w)$ of the flat equilateral triangle $ABC$ located in $z$ onto the
zero-angled triangle $ABC$ in $w$, used in the derivation of \eq{eq:ded1}, is constructed in four
sequential steps, shown in the \fig{fig:suppl01}.

First, we map the triangle $ABC$ in $z$ onto the upper half-plane $\zeta$ of auxiliary complex
plane $\zeta$ with three branching points at 0, 1 and $\infty$ -- see the \fig{fig:suppl01}a-b.
This mapping is realized by the function $z(\zeta)$:
\be
z(\zeta)=\frac{\Gamma(\frac{2}{3})}{\Gamma^2(\frac{1}{3})} \int_0^{\zeta}
\frac{d\xi}{\xi^{2/3}(1-\xi)^{2/3}}
\label{eq:z(zeta)}
\ee
with the following coincidence of branching points:
\be
\left\{
\begin{array}{lcl}
A(z=0)& \leftrightarrow &  A(\zeta=0)\medskip \\
B(z=1) & \leftrightarrow & B(\zeta=1) \medskip \\
C(z=e^{-i\frac{\pi}{3}}) & \leftrightarrow & C(\zeta=\infty)
\end{array} \right.
\label{eq:points1}
\ee

Second step consists in mapping the auxiliary upper half-plane $\Im\zeta>0$ onto the circular
triangle $ABC$ with angles $\{\alpha,\alpha,0\}$ -- the fundamental domain of the Hecke group
\cite{hecke} in $r$, where we are intersted in the specific case $\{\alpha,\alpha,0\}=\{0,0,0\}$ --
see \fig{fig:suppl01}b-c. This mapping is realized by the function $\zeta(r)$, constructed as
follows \cite{koppenfels}. Let $\zeta(r)$ be the inverse function of $r(\zeta)$ written as a
quotient
\be
r(\zeta) = \frac{\phi_1(\zeta)}{\phi_2(\zeta)}
\label{eq:quot}
\ee
where $\phi_{1,2}(\zeta)$ are the fundamental solutions of the 2nd order differential equation of
Picard-Fuchs type:
\begin{multline}
\zeta(\zeta-1) \phi''(\zeta)+\big((a+b+1)\zeta-c\big) \phi'(\zeta) \\ + ab\phi(\zeta)=0
\label{eq:fundam}
\end{multline}

Following \cite{caratheodory,koppenfels}, the function $r(\zeta)$ conformally maps the generic
circular triangle with angles $\{\alpha_0=\pi|c-1|,\alpha_1=\pi|a+b-c|, \alpha_\infty=\pi|a-b|\}$
in the upper halfplane of $w$ onto the upper halfplane of $\zeta$. Choosing $\alpha_{\infty}=0$ and
$\alpha_0=\alpha_1=\alpha$, we can express the parameters $(a,b,c)$ of the equation \eq{eq:fundam}
in terms of $\alpha$, taking into account that the triangle $ABC$ in the \fig{fig:suppl01}c is
parameterized as follows $\{\alpha_0,\alpha_1,\alpha_\infty \}=\{\alpha,\alpha,0\}$ with
$a=b=\frac{\alpha}{\pi}+\frac{1}{2}, c=\frac{\alpha}{\pi}+1$. This leads us to the following
particular form of equation \eq{eq:fundam}
\begin{multline}
\zeta(\zeta-1) \phi''(\zeta)+\Big(\frac{\alpha}{\pi}+1\Big)\big(2\zeta-1\big) \phi'(\zeta)
\\ +\Big(\frac{\alpha}{\pi} +\frac{1}{2}\Big)^2\phi(\zeta)=0
\label{eq:fundam2}
\end{multline}
where $\alpha = \frac{\pi}{m}$ and $m=3,4,...\infty$. For $\alpha=0$ Eq.\eq{eq:fundam2} takes an
especially simple form, known as Legendre hypergeometric equation \cite{golubev,hille}. The pair of
possible fundamental solutions of Legendre equation are
\be
\begin{array}{l}
\phi_1(\zeta)=F\big(\frac{1}{2},\frac{1}{2},1,\zeta\big) \medskip \\
\phi_2(\zeta)=iF\big(\frac{1}{2},\frac{1}{2},1,1-\zeta\big)
\label{eq:hyp}
\end{array}
\ee
where $F(...)$ is the hypergeometric function. From \eq{eq:quot} and \eq{eq:hyp} we get $r(\zeta) =
\frac{\phi_1(\zeta)}{\phi_2(\zeta)}$. The inverse function $\zeta(r)$ is the so-called modular
function, $k^2(r)$ (see \cite{golubev,hille,jacobi2} for details). Thus,
\be
\zeta(r) \equiv k^2(r) = \frac{\theta_2^4(0,e^{i\pi r})}{\theta_3^4(0,e^{i\pi r})}
\label{eq:zeta(r)}
\ee
where $\theta_2$ and $\theta_3$ are the elliptic Jacobi $\theta$-functions \cite{jacobi2,mum},
\be
\begin{array}{l}
\disp \theta_2\left(\chi,e^{i\pi w}\right)=2e^{i{\pi \over 4} r}
\sum_{n=0}^{\infty} e^{i\pi r n(n+1)}\cos (2n+1)\chi \medskip \\
\disp \theta_3\left(\chi,e^{i\pi r}\right)=1+2\sum_{n=1}^{\infty} e^{i\pi r n^2}\cos 2n\chi
\end{array}
\ee
and the correspondence of branching points in the mapping $\zeta(r)$ is as follows
\be
\left\{
\begin{array}{lcl}
A(\zeta=0) & \leftrightarrow & A(r=\infty) \medskip \\
B(\zeta=1) & \leftrightarrow & B(r=0) \medskip \\
C(\zeta=\infty) & \leftrightarrow & C(r=-1)
\end{array} \right.
\label{eq:points1}
\ee

Third step, realized via the function $r(w)$, consists in mapping the zero-angled triangle $ABC$ in
$r$ into the symmetric triangle $ABC$ located in the unit disc $w$ -- see \fig{fig:suppl01}c-d. The
explicit form of the function $r(w)$ is
\be
r(w) = e^{-i\pi/3}\frac{e^{2i\pi/3}-w}{1-w}-1
\label{eq:r(w)}
\ee
with the following correspondence between branching points:
\be
\left\{
\begin{array}{lcl}
A(r=\infty) & \leftrightarrow & A(w=1) \medskip \\
B(r=0) & \leftrightarrow & B(w=e^{-2\pi i/3}) \medskip \\
C(r=-1) & \leftrightarrow & C(w=e^{2\pi i/3})
\end{array} \right.
\label{eq:points1}
\ee

Collecting \eq{eq:z(zeta)}, \eq{eq:zeta(r)}, and \eq{eq:r(w)} we arrive at the following expression
for the derivative of composite function,
\be
z'(\zeta(r(w))) = z'(\zeta)\, \zeta'(r)\, r'(w)
\ee
where $'$ stands for the derivative. We have explicitly:
$$
z'(\zeta)=\frac{\Gamma(\frac{2}{3})}{\Gamma^2(\frac{1}{3})}\,
\frac{\theta_3^{16/3}(0,\zeta)}{\theta_2^{8/3}(0,\zeta)\;\theta_0^{8/3}(0,\zeta)}
$$
and
$$
\zeta'(r)|=i\pi\frac{\theta_2^4\;\theta_0^4}{\theta_3^4}; \qquad i\frac{\pi}{4}\theta_0^4=
\frac{d}{d\zeta}\ln\left(\frac{\theta_2}{\theta_3}\right)
$$
The identity
\begin{multline}
\disp \theta'_1(0,e^{i\pi\zeta})\equiv
\left.\frac{d\theta_1(\chi,e^{i\pi\zeta})}{d\chi}\right|_{\chi=0} \\ =
\pi\theta_0(\chi,e^{i\pi\zeta})\,\theta_2(\chi,e^{i\pi\zeta})\, \theta_3(\chi,e^{i\pi\zeta})
\nonumber
\end{multline}
enables us to write
\be
\left|z'(r)\right|^2=h^2 \left|\theta'_1\left(0, e^{i\pi r}\right)\right|^{8/3}
\label{3:jacobian}
\ee
where $h= \left(\frac{16}{\pi}\right)^{1/3}\frac{\Gamma(\frac{2}{3})}{\Gamma^2(\frac{1}{3})}$, and
\be
\theta_1(\chi,e^{i\pi r})=2e^{i\frac{\pi}{4} r} \sum_{n=0}^{\infty}(-1)^n e^{i\pi
n(n+1) r}\sin (2n+1)\chi \medskip \\
\label{3:ell1}
\ee
Differentiating \eq{eq:r(w)}, we get
$$
r'(w) = \frac{i\sqrt{3}}{(1-w)^2}
$$
and using this expression, we obtain the final form of the Jacobian of the composite conformal
transformation $J(z(\zeta(r(w))))$:
\be
J(z(w)) = |z'(w)|^2 = 3h^2\frac{|\eta(r(w))|^8}{|1-w|^4}
\label{eq:composite}
\ee
where
$$
\eta(r) = \big(\theta_1'(0,e^{i\pi r})\big)^{1/3}
$$
is the Dedekind $\eta$-function (see \eq{eq:ded2}), and the function $r(w)$ is defined in
\eq{eq:r(w)}.

\end{appendix}

\end{document}